\documentclass[aps,prl,reprint,superscriptaddress,floatfix]{revtex4-2}

\newcommand{\bibo}{{\textalpha}-{BiB$_{3}$O$_{6}$}}
\newcommand{\Gru}{Gr{\"u}neisen}
\newcommand{\mode}{\omega_{n,\mathbf{k}}}
\newcommand{\mgp}{\gamma_{ij,n,\mathbf{k}}}

\usepackage{textgreek}
\usepackage{bm}
\usepackage{amsmath}
\usepackage{graphicx}
\usepackage{physics}

\begin{document}

%\preprint{}

\title{Anomalous thermal expansion and chiral phonons in BiB$_{3}$O$_{6}$}


\author{Carl P. Romao}
\email[carl.romao@mnf.uni-tuebingen.de]{}
%\thanks{}
\affiliation{Department of Chemistry, University of Oxford, Inorganic Chemistry Laboratory, South Parks Road, Oxford OX1 3QR, UK}
\affiliation{Section for Solid State and Theoretical Inorganic Chemistry, Institute of Inorganic Chemistry, University of T{\"u}bingen, Auf der Morgenstelle 18, D-72076 T{\"u}bingen, Germany}


\date{\today}

\begin{abstract}
The origins of anomalous thermal expansion in the chiral monoclinic solid {\bibo} have been studied through \emph{ab initio} calculations. Positive and negative axial thermal expansion are shown to be driven by librations of borate units, elastic anisotropy, and most notably by chiral acoustic phonons involving elliptical revolutions of bismuth atoms. The chirality of the lattice gives rise to these modes by allowing the transverse acoustic branches to have opposite circular polarizations, only one of which couples strongly to the lattice strains. These results further understanding of relationships between crystallographic symmetry and physical properties.
\end{abstract}

%\keywords{}

\maketitle

\begin{figure}[h]
\centering
  \includegraphics[width=8.6cm]{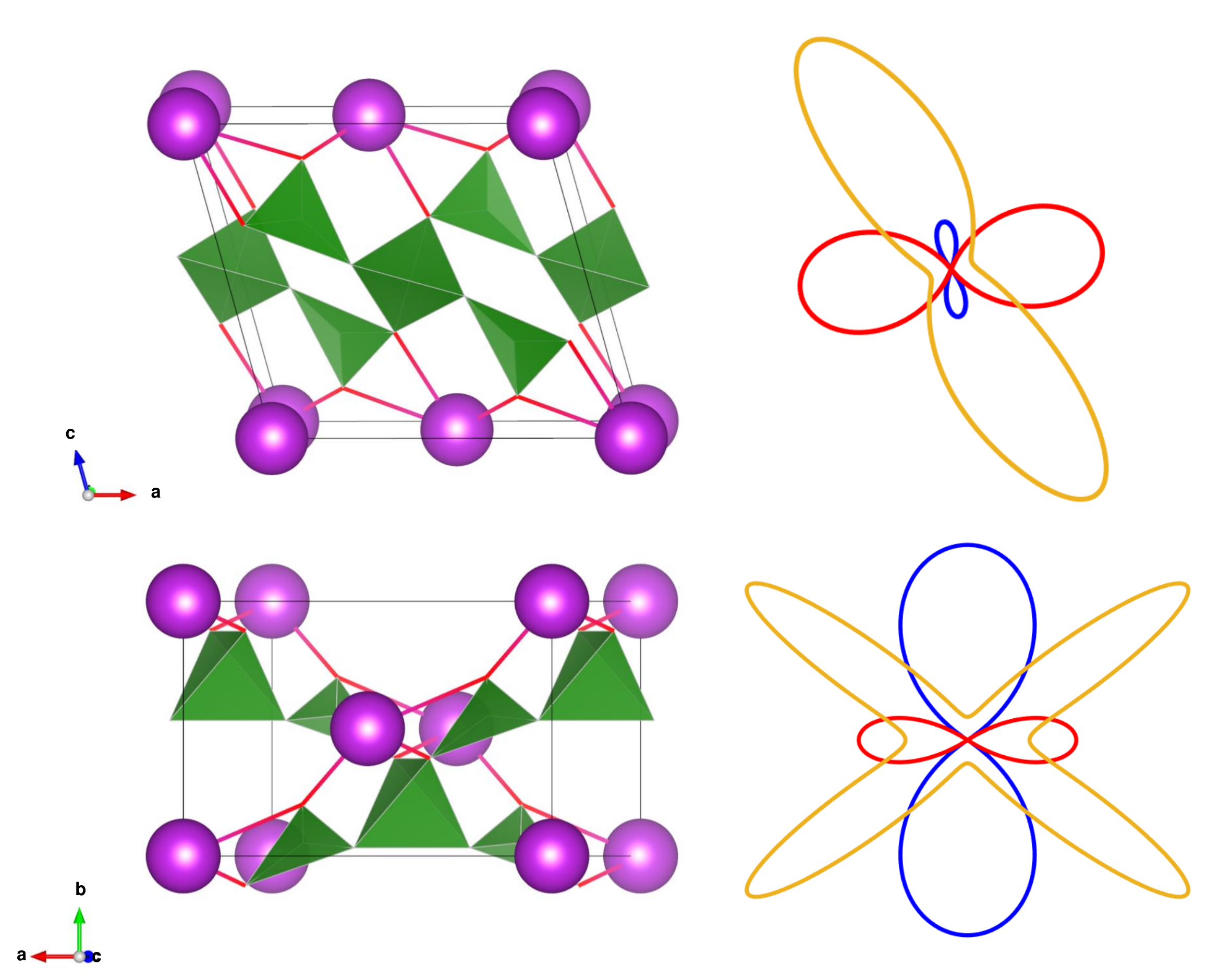}
  \caption{At left, the structure of {\bibo} is shown in the $\mathbf{ac}$ plane (top) and the $\mathbf{ab}$ plane (bottom), with bismuth atoms coloured in magenta and borate units coloured in green \citep{stein2007temperature}. At right, the indicatrix of directional Young's modulus is shown in orange, and those of positive and negative thermal expansion are shown in blue and red, respectively \citep{teng2002anisotropic, kang2015negative}.}
  \label{fgr:structure}
\end{figure}

The monoclinic {\textalpha} phase of bismuth borate (BiB$_{3}$O$_{6}$), in addition to having promising applications in nonlinear optics \citep{hellwig1998exceptional, pinel2012generation} and quantum computing \citep{chen2017observation, wang2016experimental}, possesses the unusual physical properties of uniaxial negative thermal expansion (NTE) \citep{teng2002anisotropic, stein2007temperature} and negative linear compressibility (NLC) \citep{kang2015negative, cairns2015negative} (Fig. \ref{fgr:structure}). The chiral $C2$ space group adopted by {\bibo} \citep{stein2007temperature} has unusually low symmetry for an NTE material \citep{Romao_2013, dove2016negative}, or indeed more generally for an inorganic solid. The chirality of {\bibo} permits it to have chiral phonons, \emph{i.e.}, phonons with a circular polarization and pseudoangular momentum \citep{zhang2015chiral, gao2018nondegenerate, zhu2018observation}. Phonon chirality has been shown to affect the anharmonic properties of phonons through the phonon Hall effect \citep{zhang2015chiral}, and through modification of symmetry constraints for scattering \citep{pandey2018symmetry}, but these previous studies were limited to 1D and 2D materials. The effects of phonon chirality in a 3D crystal are studied here through their relationship to the related anharmonic property of thermal expansion \citep{kennedy2005unusual}.

The thermal expansion of {\bibo} is highly anisotropic, with large NTE along $\mathbf{a}$ and large positive thermal expansion along $\mathbf{b}$ \citep{teng2002anisotropic, stein2007temperature}. Such behaviour is characteristic of flexible framework materials, \emph{i.e.}, materials which have a mixture of compliant and stiff directions. Indeed, {\bibo} is also very elastically anisotropic: in the $\mathbf{ac}$ plane the ratio of Young's moduli between the stiffest and most compliant direction ($Y_{\text{max}}/Y_{\text{min}}$) is $25$ (Fig. \ref{fgr:structure}, top); in the $\mathbf{ab}$ plane $Y_{\text{max}}/Y_{\text{min}} = 15$ (Fig. \ref{fgr:structure}, bottom) \citep{teng2002anisotropic}. By these standards, {\bibo} is an extraordinarily flexible inorganic material; it can be compared to the colossal NTE material Ag$_3$[Co(CN)$_6$], for which $Y_{\text{max}}/Y_{\text{min}} = 6.0$ \citep{wang2016correlation, goodwin2008colossal}, and to ZnAu$_2$(CN)$_4$, which exhibits extreme NLC and has $Y_{\text{max}}/Y_{\text{min}} = 3.8$ \citep{gupta2017anomalous}.

Axial NTE has been studied extensively in the tetragonal, orthorhombic, and hexagonal crystal families due to the appearance of anomalously large-magnitude thermal expansion \citep{goodwin2008colossal}, connections with ferroelectricity \citep{senn2015negative, ritz2018interplay}, and the use of chemical control to achieve zero thermal expansion \citep{romao2015zero}. However, NTE is extremely rare in monoclinic and triclinic crystals \citep{Romao_2013, dove2016negative}. Outside of \bibo, the phenomenon has only been studied in molecular crystals, where the mechanism commonly involves the sliding of layers relative to each other \citep{bhattacharya2012uniaxial, saha2017thermal}. Elastic interactions between layers in {\bibo} are strong in the direction coinciding with Bi--O bonds, and thermal expansion has a small positive value in that direction (Fig. \ref{fgr:structure}, top); {\bibo} therefore offers a unique opportunity to study NTE in a chiral monoclinic 3D framework solid. 

Herein is reported a theoretical investigation of the origins of anomalous thermal expansion in {\bibo}, and of the influence of phonon chirality on this bulk property. The contribution of individual phonons to thermal expansion was assessed following the method of Ref. \citenum{romao2017anisotropic}; \emph{i.e.,} by using uniaxial stress perturbations to calculate mode {\Gru} parameters ($\mgp$):
\begin{align}
\label{yunidef}
{\gamma}_{ij,n,\mathbf{k}} &= - \frac{1}{s_{ijij}} \left( \frac{\partial ~\text{ln}~ \mode}{\partial \sigma_{ij}} \right)_{T,\sigma'},
\end{align}
where $\mathbf{s}$ is the elastic compliance tensor. These mode {\Gru} parameters were averaged, weighted by their contribution to the heat capacity ($C_{\bm{e}}$), to give bulk {\Gru} parameters ($\gamma_{ij}$), and subsequently determined directional coefficients of thermal expansion (CTEs, $\alpha_{ij}$) as \citep{romao2017anisotropic}: 
\begin{align}
\label{simple}
\alpha_{ij} = s_{ijij} {\gamma}_{ij} \frac{C_{\bm{e}}}{V}.
\end{align}
This recently developed method relates each mode {\Gru} parameter to the thermal expansion in a single direction, and therefore allows identification of the phonons which drive thermal expansion along each principal axis \citep{romao2017anisotropic}.

Phonon band structures and the elastic tensor were calculated using density functional theory (DFT) \citep{gonze1997dynamical,van2016interatomic} within the software package \textsc{Abinit} \citep{Gonze2016106, bottin2008large, bjorkman2011cif2cell, torrent2008implementation}; example input files are available as part of the Supplemental Material \citep{si}. The uniaxial stress perturbations used were $\sigma_{11}$, $\sigma_{22}$, $\sigma_{33}$, and $\sigma_{13}$; the Cartesian axes $\mathbf{x}$ and $\mathbf{y}$ coincide with the lattice vectors $\mathbf{a}$ and $\mathbf{b}$, respectively, and the $\mathbf{z}$ axis is therefore the stacking direction of the borate layers (see Fig. \ref{fgr:structure}). DFT methods were validated by comparison of calculated elastic and thermal expansion tensors to experimental values \citep{kang2015negative, teng2002anisotropic}. As described in the Supplemental Material \citep{si}, several approaches were trialled \citep{kohn1965self, he2014accuracy, grimme2010consistent, becke2006simple, perdew1996generalized, hamann2013optimized, monkhorst1976special, abipsps, psps}. The most accurate results were found for a dispersion-corrected exchange--correlation functional \citep{grimme2010consistent, becke2006simple}; the calculations underestimated the magnitude of thermal expansion somewhat but were able to qualitatively reproduce the experimental behaviour \citep{si}.

\begin{figure}[h]
\centering
  \includegraphics[width=8.6cm]{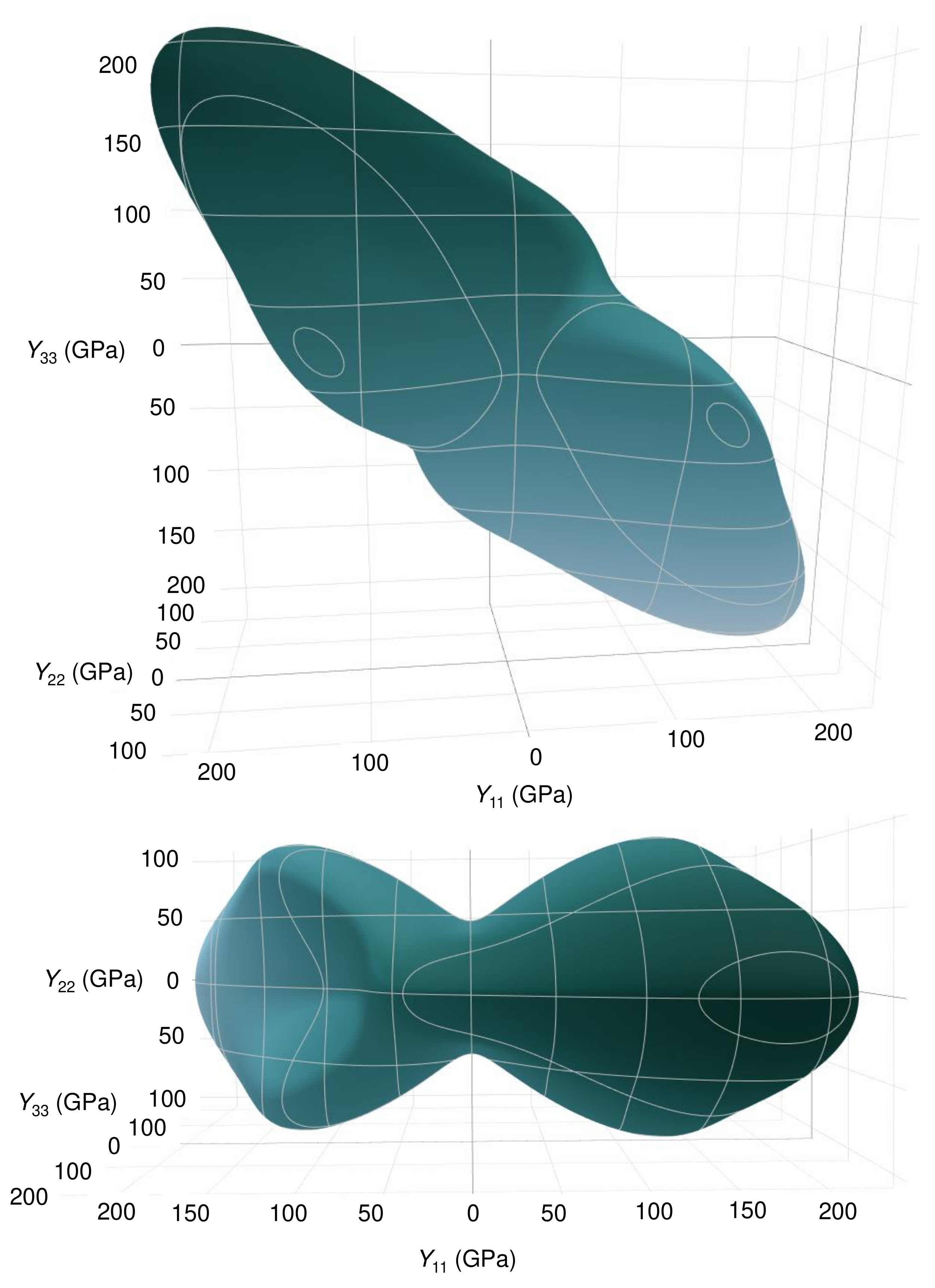}
  \caption{Calculated directional Young's modulus ($Y_{ii}$) of {\bibo}, shown as a blue surface. The views of the $\mathbf{ac}$ (top) and $\mathbf{ab}$ (bottom) planes correspond to Fig. \ref{fgr:structure}. Visualization generated with \textsc{Elate} \citep{gaillac2016elate}.}
  \label{fgr:y}
\end{figure}

\begin{figure*}[ht]
\centering
  \includegraphics[width=18cm]{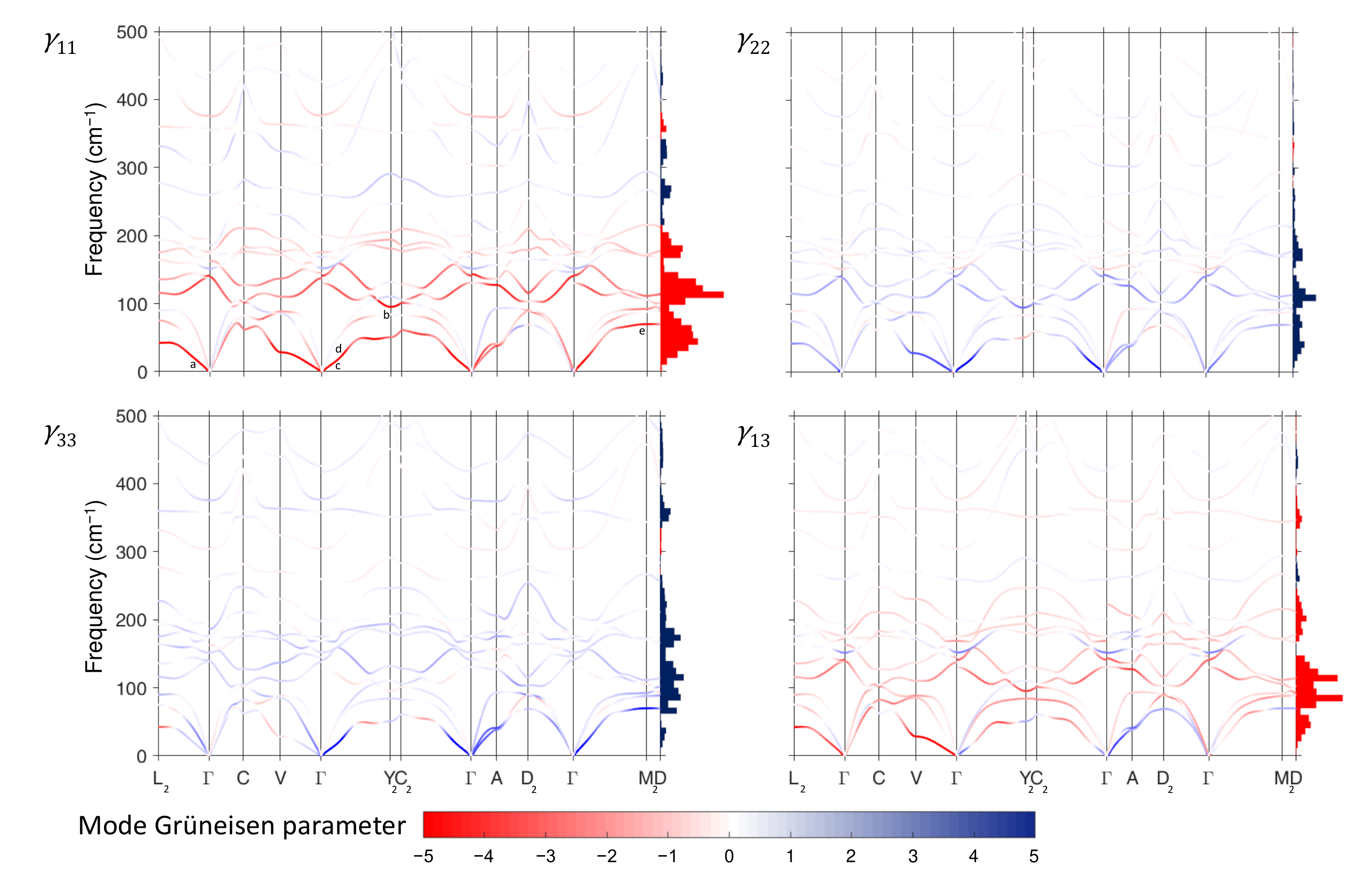}
  \caption{Phonon band structure of {\bibo}, with bands coloured according to their directional mode {\Gru} parameters ($\mgp$), calculated using the method of Ref. \citenum{romao2017anisotropic}. Phonons with energies greater than 500 cm$^{-1}$ do not contribute significantly to thermal expansion and are not shown. The density of states ($\rho$), weighted by the {\Gru} parameters as $\sum_{\mathbf{k}} \rho_{\mathbf{k}}(\omega) \gamma_{ij,\mathbf{k}}(\omega)$, is shown as a histogram at the right of each plot, with positive values coloured in blue and negative values in red. Special points in the Brillouin zone were selected following Ref. \citenum{hinuma2017band}. The modes marked a--e are visualized in Fig. \ref{fgr:phon}.}
  \label{fgr:gru}
\end{figure*}

The calculated directional Young's moduli ($Y_{ii} = 1/s_{iiii}$) are shown in Fig. \ref{fgr:y}, and phonon band structures, coloured by their directional mode {\Gru} parameters ($\mgp$), are shown in Fig. \ref{fgr:gru}. Thermal expansion results from the anharmonicity of both acoustic bands and low-energy optic bands, predominantly those with energies below 200 cm$^{-1}$.  The small mode {\Gru} parameters for perturbations along $\mathbf{b}$ ($\gamma_{22}$) indicate that the large thermal expansion in this direction is driven by elastic anisotropy, as the directional thermal expansion is inversely proportional to the directional Young's modulus (Eq. \ref{simple}) and the directional Young's modulus is minimal along $\mathbf{b}$  (Fig. \ref{fgr:y}). Indeed, in the $\mathbf{ab}$ plane the directional Young's modulus of {\bibo} exhibits the bowtie shape characteristic of flexible framework materials with highly anisotropic thermal expansion, with maxima in $Y_{ii}$ close to minima in $|\alpha_{ii}|$ (Fig. \ref{fgr:structure}) \citep{romao2017anisotropic, ortiz2012anisotropic, ortiz2013metal, dolabdjian2018synthesis}.

%Indeed, inspection of the directional Young's modulus (Fig. \ref{fgr:y}) indicates that, despite its layered structure, {\bibo} behaves elastically like a flexible framework: it has a mixture of compliant directions ([1 0 0], [0 1 0], and [1 0 $-$1]) and stiff directions ([1 1 0], [$-$1 1 0], and [0 0 1]). 

%While understanding of the origins of NTE has often focused on the low-energy optic phonons corresponding to rigid unit modes, acoustic modes are often also significant contributors \citep{rimmer2014acoustic, fang2014common, dove2016negative}, as is seen here.

\begin{figure}[h]
\centering
  \includegraphics[width=8.6cm]{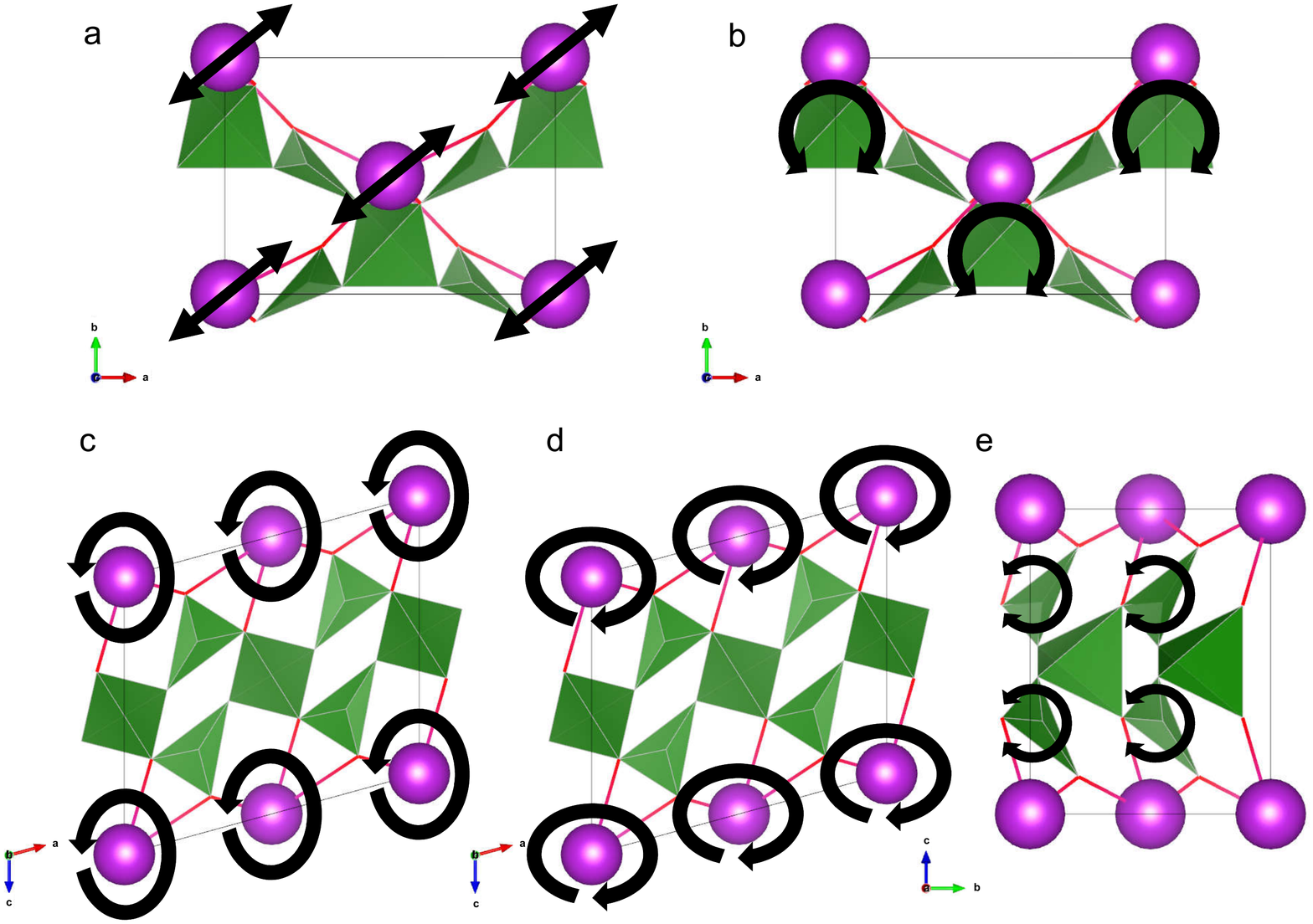}
  \caption{Cartoon view of the eigenvectors of several phonons in {\bibo}. a: Linear oscillations of Bi atoms; b: librations of BO$_4$ tetrahedra; c and d:  elliptical orbits of Bi atoms about their average positions; e: librations of BO$_3$ triangles (see Fig. \ref{fgr:gru} and Fig. \ref{fgr:poly} for mode wavevectors and energies). In order to allow these modes to be viewed directly, animations have been provided in the Supplemental Material \citep{si}.}
  \label{fgr:phon}
\end{figure}

The origins of anomalous thermal expansion in {\bibo} can be understood further by inspection of the eigenvectors of modes with large mode {\Gru} parameters. The complete set of eigenvectors are available in the Supplemental Material \citep{si}, in a format which allows their visualization in animated form \cite{phonweb}. The eigenvectors show that thermal expansion in {\bibo} is driven primarily by two types of low-energy phonons: those which involve large displacements of the Bi atoms (\emph{e.g.}, Fig. \ref{fgr:phon} a, c and d), and those which primarily involve motions of rigid BO$_{3}$ triangles and BO$_{4}$ tetrahedra (\emph{e.g.}, Fig. \ref{fgr:phon} b and e). The modes in this latter group are rocking motions of O atoms in the B--O--B bonds (Fig. \ref{fgr:phon} c) and the Bi--O--B bonds (Fig. \ref{fgr:phon} d). Such librations of bridging atoms are known to be a common mechanism for NTE in flexible framework materials \citep{Romao_2013, dove2016negative}. 

The low energy acoustic phonons which drive positive thermal expansion along $\mathbf{b}$, and contribute significantly to negative thermal expansion along $\mathbf{a}$, involve large displacements of the Bi atoms as well as smaller transverse motions of the borate framework. These phonons can be divided into two categories; the first being modes where the Bi atom oscillates along a linear trajectory (\emph{e.g.} Fig. \ref{fgr:phon} a). The second category is comprised of modes where the Bi atoms orbit their average positions (\emph{e.g.} Fig. \ref{fgr:phon} c). These orbits are elliptical due to the anisotropic coordination environment of the Bi atoms (see Fig. \ref{fgr:structure}). Acoustic modes have been found to contribute to NTE in 1D materials \citep{fang2014common, dove2016negative} and in the metal-organic framework MOF-5 \citep{rimmer2014acoustic}; although the mechanism in those cases simply involves sinusoidal oscillation of chains or layers, respectively. In {\bibo}, these transverse displacements are present in the borate layers, and they are strongly coupled to the revolutions of the Bi atoms. The elliptical revolution of the Bi atom shown in Fig. \ref{fgr:phon} c is associated with a larger transverse displacement of the borate layer than the opposing revolution shown in Fig. \ref{fgr:phon} d. These modes are chiral; the coordinated rotation of the Bi atoms has a distinct handedness and a circular polarization results. 

The circular polarization ($\mathbf{S}$) of a phonon can be quantified by performing a series of basis transformations to its eigenvectors ($\ket{\epsilon}$) so that each sublattice has a right- or left-handed circular polarization with respect to a given axis \citep{zhang2015chiral}. This method has previously been applied to one- and two-dimensional materials \citep{zhang2015chiral, gao2018nondegenerate, pandey2018symmetry}; to examine phonon chirality in {\bibo} it was generalized by determining the polarization along each Cartesian axis independently. The new bases are defined, for example for a polarization along $\mathbf{b}$, as: 
$\ket{R_{1,22}} \equiv \frac{1}{\sqrt{2}} (i\ 0\ 1\ \cdot\cdot\cdot\ 0)^{\text{T}}$, $\ket{L_{1,22}} \equiv \frac{1}{\sqrt{2}} (-i\ 0\ 1\ \cdot\cdot\cdot\ 0)^{\text{T}}$, $\ket{R_{n,22}} \equiv \frac{1}{\sqrt{2}} (0\ \cdot\cdot\cdot\ i\ 0\ 1)^{\text{T}}$, and $\ket{L_{n,22}} \equiv \frac{1}{\sqrt{2}} (0\ \cdot\cdot\cdot\ -i\ 0\ 1)^{\text{T}}$, where $n$ is the number of atoms \citep{zhang2015chiral, gao2018nondegenerate}. The phonon polarization along $\mathbf{b}$ ($S_{22}$) is then the sum of the polarizations of each sublattice:
\begin{align}
S_{22} = \sum_{\alpha=1}^{n} (\abs{\bra{R_{\alpha,22}}\ket{\epsilon}}^2-\abs{\bra{L_{\alpha,22}}\ket{\epsilon}}^2)\hbar;
\end{align}
the circular polarizations $S_{11}$ and $S_{33}$ are obtained by rotation of the basis vectors \citep{zhang2015chiral, gao2018nondegenerate}.

The circular polarizations along $\mathbf{y}$ of the low-energy phonons are shown in Fig. \ref{fgr:poly} (the polarizations along $\mathbf{x}$ and $\mathbf{z}$ are shown in the Supplemental Material \citep{si}). Comparison of Fig. \ref{fgr:gru} and Fig. \ref{fgr:poly} shows that phonon chirality can significantly influence the mode {\Gru} parameters. The transverse acoustic branches can have opposite chirality (for example along \textGamma $\rightarrow$ C$_{2}$), with one branch involving right-handed revolutions of Bi atoms (Fig. \ref{fgr:phon} c) and the other having left-handed revolutions (Fig. \ref{fgr:phon} d). The right-handed branch couples strongly to the lattice strains, as evidenced by its large mode {\Gru} parameters, while the left-handed branch does not couple significantly.  Note that the two branches do not simply mirror each other; branches with the same energies but reversed chirality exist along paths related by time-reversal symmetry, \emph{e.g.} \textGamma $\rightarrow$ C$_{2}^\prime$ \citep{zhang2015chiral, hinuma2017band}. Specific signs of the phonon polarization do not lead to specific mode {\Gru} parameters; the chirality is important because the polarized modes correspond to motions which would be symmetry-forbidden for phonon eigenvectors in an achiral lattice.

\begin{figure}[h]
\centering
  \includegraphics[width=9cm]{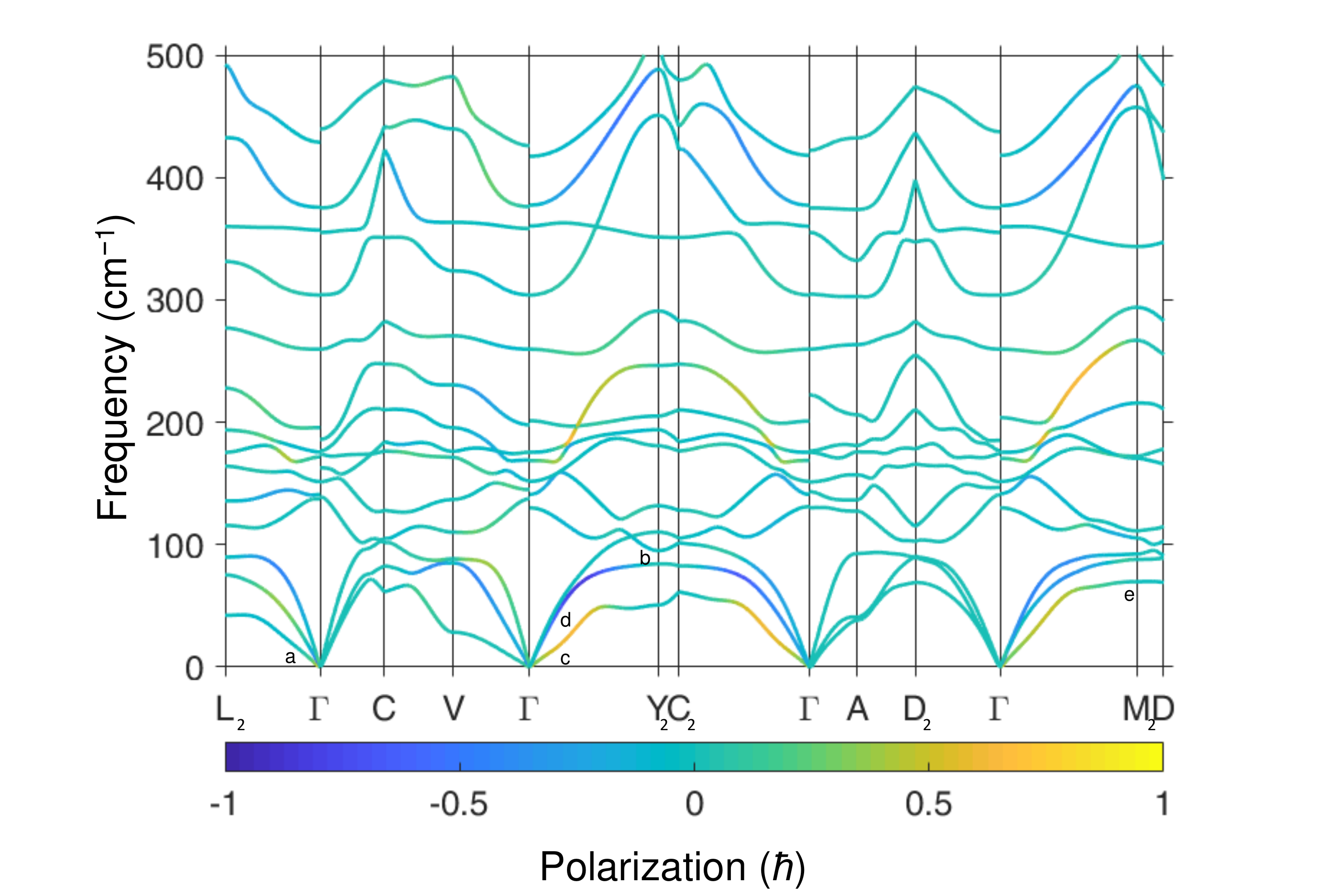}
  \caption{Phonon band structure of {\bibo}, with bands coloured according to their circular polarizations along $\mathbf{b}$ ($S_{22}$). A polarization of $\hbar$ corresponds to a fully right-polarized phonon \citep{zhang2015chiral}. Special points in the Brillouin zone were selected following Ref. \citenum{hinuma2017band}. The modes marked a--e are visualized in Fig. \ref{fgr:phon}.}
  \label{fgr:poly}
\end{figure}

Fig. \ref{fgr:gru} and Fig. \ref{fgr:poly} show that the acoustic modes with circular polarization along $\mathbf{b}$ are significant contributors to the anomalous thermal expansion of {\bibo}, as they have large values of $\gamma_{11}$, $\gamma_{22}$, and $\gamma_{33}$. These mode {\Gru} parameters are strongly influenced by dispersive interactions, as their signs and magnitudes vary widely between calculations including a dispersion correction (Fig. \ref{fgr:gru}) and those performed without it (see the Supplemental Material \citep{si}). This discrepancy suggests that the qualitative level of accuracy in the calculated CTE could be due to underestimation of the {\Gru} parameters of the chiral modes.

The relationship between phonon chirality and thermal expansion in {\bibo} can be understood by examination of the chiral acoustic modes. If the structure of {\bibo} were achiral,  adjacent Bi atoms would be required to revolve in opposite directions in order to maintain zero circular polarization \citep{zhang2015chiral, gao2018nondegenerate}. Therefore, modes which involve such revolutions are limited to the edges of the Brillouin zone in achiral crystals. As Fig. \ref{fgr:gru} shows, the modes that contribute most strongly to thermal expansion lie away from the edges of the Brillouin zone. Therefore, revolutions of Bi atoms with small wavevectors couple strongly to the lattice strains, and if such phonons were disallowed by symmetry, the thermal expansion of {\bibo} would be less extreme. Chirality therefore acts like a degree of freedom, or as a kind of flexibility: it allows a specific type of low-energy phonon in {\bibo}, analogously to how corner-linked topologies of coordination polyhedra can allow low-energy rigid unit modes (RUMs) to exist \citep{dove2016negative}. Since the chiral phonons can have negative mode {\Gru} parameters, it is possible for them to act as soft modes and cause pressure-induced phase transitions \citep{dove2016negative, dinnebier2009high}. 

The chiral nature of {\bibo} is quite unusual among inorganic framework materials, and here the connection between that unusual structural feature and the phonons which cause anomalous thermal expansion has been revealed. It is interesting to note that, despite the general rarity of chiral inorganic crystals, several very prominent NTE materials crystallize in chiral space groups: {\textalpha}-ZrW$_{2}$O$_{8}$, which was the first material discovered to have isotropic NTE over a broad temperature range, crystallizes in $P 2_{1} 3$ \citep{mary1996negative}, and {\textbeta}-eucryptite, which is widely used in zero-thermal-expansion glass ceramics, crystallizes in $P 6_{4} 2 2 2$ \citep{lichtenstein1998anisotropic}. The structure of {\textbeta}-eucryptite is identical to that of {\textbeta}-quartz, which also displays NTE \citep{welche1998negative}. 

The chiral modes found in {\bibo} are distinct from the RUMs which have been used successfully to understand NTE in some materials \citep{dove2016negative}. However, the RUM model has been found to not apply to many NTE materials, either because of overconstrained coordination polyhedra \citep{rimmer2015simulation}, or because the polyhedra simply are not rigid \citep{rimmer2015negative, baise2018negative}. In materials where the RUM model fails, the origins of NTE often remain incompletely understood, and it is possible that, for chiral systems, calculations of mode circular polarizations could bring additional clarity.

In {\textalpha}-ZrW$_{2}$O$_{8}$, NTE is caused by a wide range of low-energy modes involving motions of rigid WO$_4$ tetrahedra and Zr--O bonds \citep{rimmer2015negative, gupta2013negative}. These modes can be described as fluctuations towards denser structures where non-bridging O atoms on WO$_4$ tetrahedra form W--O--W chains \citep{baise2018negative, evans1997compressibility}. The rigidity constraints could allow chiral motions of the Zr atoms and of the non-bridging O atoms on the WO$_4$ tetrahedra. For example, a revolution of a terminal O atom would bring it closer in turn to each of the three equivalent positions where it could form a W--O--W bond \citep{baise2018negative}. As the terminal O atoms are permitted by network topology and crystallographic symmetry to undergo chiral revolutions, it would not be surprising to find such modes with the methods used here. 

In conclusion, two types of phonon were found to contribute significantly to thermal expansion in {\bibo}: motions of the rigid borate units, and modes involving oscillations and revolutions of the bismuth atoms. These revolutions correspond to chiral acoustic phonons; along certain directions in reciprocal space the two transverse acoustic branches have opposite circular polarizations, with only one coupling strongly to the lattice strains. Were the lattice achiral, such modes would be disallowed by symmetry except at the edges of the Brillouin zone, where they have small mode {\Gru} parameters. Phonon chirality has therefore been related to the macroscopic behaviour of a three-dimensional material; demonstrating that structural chirality permits specific types of modes in a manner analogous to structural flexibility. The resulting chrial phonons have low energies and large mode {\Gru} parameters; they thereby drive thermal expansion and potentially could act as soft modes and cause phase transitions. This result reveals a connection between the absence of a particular type of crystallographic symmetry and an emergent physical property, and can be applied to the study of structure--property relationships in other chiral crystals.

The author acknowledges the support of the National Sciences and Engineering Council of Canada. Computational resources were provided by the University of Oxford, Department of Chemistry, and the UK's HEC Materials Chemistry Consortium, which is funded by EPSRC (EP/L000202). This work used the ARCHER UK National Supercomputing Service (http://www.archer.ac.uk).

\bibliography{bibo}

\end{document}